\documentstyle[psfig,floats,epsf,prl,aps,pstricks,psfig,pst-plot,rotate]{revtex}

\begin{document}
\twocolumn[\hsize\textwidth\columnwidth\hsize\csname
@twocolumnfalse\endcsname

\title{
{\rm\small\hfill Phys. Rev. Lett., in press.}\\
Adsorption of Xe atoms on metal surfaces: New insights from 
first-principles calculations}
\author{Juarez L. F. Da Silva$^{1}$, Catherine Stampfl$^{1,2}$, and Matthias 
Scheffler$^{1}$}
\address{$^{1}$Fritz-Haber-Institut der Max-Planck-Gesellschaft, Faradayweg 
4-6, D-14195 Berlin-Dahlem, Germany \\ $^{2}$Department of Physics and 
Astronomy, Northwestern University, Evanston, IL 60208, USA }
\date{\today}
\maketitle

\begin{abstract}

The adsorption of rare gases on metal surfaces serve as the paradigm of weak 
adsorption where it is typically assumed that the adsorbate occupies 
maximally coordinated hollow sites. Density-functional theory calculations 
using the full-potential linearized augmented plane wave method for Xe 
adatoms on Mg(0001), Al(111), Ti(0001), Cu(111), Pd(111), and Pt(111), show, 
however, that Xe prefers low-coordination on-top sites in {\em all} cases. We 
identify the importance of polarization and
a site-dependent Pauli repulsion in actuating 
the site preference and the principle nature of the 
rare-gas atom--metal surface interaction.
 {\em Copyright (2002) by The American Physical Society.}
\vspace{0.2cm}


\end{abstract}
\vskip2pc]

The adsorption of rare-gas atoms on metal surfaces represent prototypical 
physical adsorption systems~\cite{Bruch-1997,Zeppenfeld-LB}. Noting that 
these adatoms have closed shells, it is typically assumed that the 
adatom-substrate interaction is determined by an interplay of van der Waals 
attraction and Pauli repulsion, and that these adsorbates prefer maximally 
coordinated sites~\cite{Bruch-1997}. However, the concept of the van der 
Waals interaction (dispersion forces) is valid for large distances where 
orbitals almost do not overlap, and at closer distances, near the equilibrium 
geometry of an adsorbate on a surface, the {\em direct} interaction of 
adsorbate and substrate orbitals can be significant. Thus, although the van 
der Waals {\em tail} is not described by state-of-the-art 
exchange-correlation functionals, the short-range attraction, as well as the 
Pauli repulsion, is treated properly. This appears to be valid in particular 
for the adsorption of heavier rare-gas atoms, but also for the {\em 
scattering} of He and Ne atoms at metal surfaces, the interatomic distances 
{\em at the turning point} are sufficiently small that density-functional 
theory (DFT) calculations provide the correct description~\cite{Petersen-1996}.

With regard to the preferred adsorption site, about ten years ago it was 
argued that Xe on Pt(111) adsorbs in low-coordination on-top sites instead of 
the expected high-coordination hollow sites~\cite{Gottlieb-1990}; a quite 
surprising, not widely accepted, and certainly not understood suggestion. 
This assessment was based on He-atom diffraction data from the 
low-temperature ($T \le 60$~K) uniaxially compressed phase of Xe on 
Pt(111)~\cite{Kern-1987,Zeppenfeld-1988}. Later,  a spin-polarized 
low-energy electron diffraction (SP-LEED) study of Xe-adsorption ``showed'' 
that Xe atoms adsorb preferentially in hollow sites on 
Pt(111)~\cite{Potthoff-1995} and Pd(111)~\cite{Hilgers-1995} in the 
$(\sqrt{3} \times \sqrt{3})R30^{\circ}$ structure (from now on labeled 
$\sqrt{3}$). These latter findings were subsequently contradicted by
recent LEED studies, which reported that Xe in the $\sqrt{3}$ structure 
on Cu(111), Pd(111), and Pt(111) adsorbs in the on-top 
sites~\cite{Seyller-1998}. 

Thus, despite the conceived ``simplicity'' of these adsorption systems, even 
the most basic questions have not been answered satisfactorily: What is the 
adsorption site and {\em why} is it what it is? and what is the nature of the 
adsorbate-surface bond?

With respect to first-principles studies, the first good-quality calculation 
was performed by M\"{u}ller~\cite{Muller-1990}, who studied Xe on Pt(111) 
using the cluster approach and the local-density approximation (LDA). It was 
concluded that the on-top site was energetically preferred over the hollow 
site by 30~meV. The interaction was described in terms of a hybridization of 
Xe $5p$ orbitals with metal $d$-states, with the on-top site affording the 
greater mixing and hence being the preferred adsorption site. 
Some questions about this work were 
raised (e.g. \cite{Seyller-1998}): $(i)$ Besides the general concern about 
the validity of the cluster approach, such calculations may only relate to 
low coverages, and not to the case of Xe in the $\sqrt{3}$ adlayer, which is 
the only structure for which good experimental data exists. $(ii)$ It is also 
unclear {\em why} the top site affords the greater mixing. As it stands, this 
is more a statement than an explanation. $(iii)$ Because on-top site 
adsorption was also found for Xe on Cu(111), for which the $d$-band lies well 
below the Fermi level, 
a crucial role of substrate $d$-states seems 
unlikely~\cite{Seyller-1998}. A recent work by Betancourt and 
Bird~\cite{Betancourt-2000} find that Xe also prefers the on-top site in the 
$\sqrt{3}$ adlayer structure on Pt(111).  And for an artificial 
$c(2 \times 2)$-Xe/Ag(001) structure, Clarke {\em et al.}~\cite{Clarke-2001} 
reported a theoretical study which also favors the on-top site. However, 
neither Betancourt and Bird, nor Clarke {\em et al.} presented a consistent 
explanation for the on-top site preference.
Finally, a recent study 
of the Xe/Cu(111) system using a ten-atom cluster model,
has investigated the interface dipole and work-function,
but the site preference was not addressed~\cite{bagus}.

To determine the adsorption site preference of Xe adatoms on metal surfaces, 
to {\em explain} the site preference, and to provide an improved 
understanding of rare-gas adsorption on metal surfaces in general, we 
performed DFT calculations using the all-electron, full-potential linearized
augmented plane wave (FP-LAPW) method~\cite{Blaha-1990,Petersen-2000} for Xe 
adatoms on the Mg(0001), Al(111), Ti(0001), Cu(111), Pd(111), and Pt(111) 
surfaces. We employed two different 
exchange-correlation functionals, namely, the LDA~\cite{Perdew-1992} and the 
generalized gradient approximation (GGA)~\cite{Perdew-1996}. 
The surfaces are treated by the 
supercell approach using the repeated slab geometries (five- and six-layer 
thick slabs separated by a vacuum region of 18~{\AA}). To determine the 
potential-energy surface (PES) of Xe on the metal surfaces we chose six sites 
on the fcc(111) surface: on-top, bridge, fcc, hcp, the midpoint between 
on-top and fcc, and the midpoint between on-top and hcp (see inset of 
Fig.~1). For each site the height of the Xe atom and the position of the 
topmost two substrate layers are fully relaxed. 

\begin{figure}[t]
 \pspicture(5,7.5)
 \rput[c](4.3,3.8){
 \psfig{figure=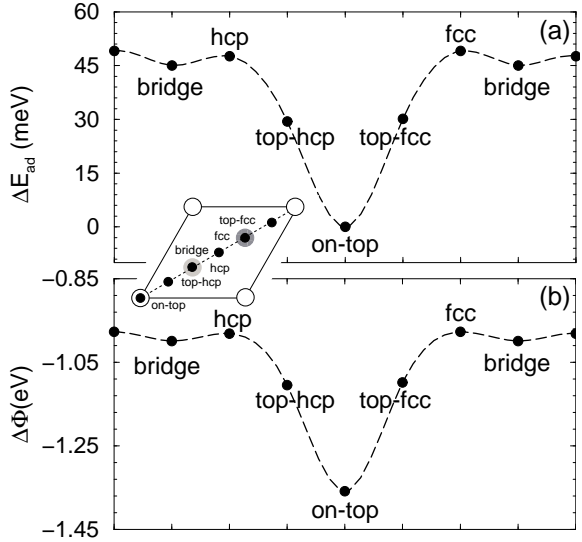,height=7.5cm}}
 \endpspicture
\caption{
DFT-LDA results for: (a) Relative adsorption energy, $\Delta E_{\rm ad}$, of 
Xe on Pt(111) (the energy of the on-top site 
is the energy zero). The inset shows a top view of the 
fcc(111)-($1 \times 1$) surface unit cell, indicating the six different sites 
considered for Xe-adsorption. Full circles denote the sites for the Xe atoms, 
the large open, pale, and dark gray circles denote the metal atoms in 
the first, second, and third layers, respectively. (b) Work-function change, 
$\Delta \Phi$. The calculated 
value for the clean surface is $\Phi^{\rm Pt(111)} = 6.06$~eV. }
\label{sites}
\end{figure}             

\begin{table}[t]
\caption{Adsorption energy of Xe in the on-top site, $E_{\rm ad}$; 
adsorption-energy difference between on-top and fcc sites, 
$E_{\rm ad}^{\rm ft}$ 
$=$ $E_{\rm ad}^{\rm fcc}$ $-$ $E_{\rm ad}^{\rm on-top}$ relative
to the on-top site; work-function 
change, $\Delta\Phi$; vertical distance of 
Xe to the first metal layer, $d_{\rm Xe-metal}$ (given as the 
average distance for the topmost layer), and substrate rumpling 
(a positive sign indicates an outward and a negative sign an 
inward displacement). In parenthesis are experimental 
results~\protect\cite{Bruch-1997,Seyller-1998}.
}
\label{CuPtPd-table}
\begin{tabular}{c|cc|cc}
          &  $E_{\rm ad}$    &$\Delta E_{\rm ad}^{\rm ft}$&\multicolumn{2}{c}{    $\Delta \Phi$~(eV)     }   \\
          &     (meV)        &     (meV)                  &     on-top        &      fcc                     \\  \hline\hline
 Mg(0001) &  $-130$~~~~~~~~  &    $~4.34$                 &     $-0.22$       &     $-0.16$                  \\
  Al(111) &  $-176$~~~~~~~~  &    $~1.12$                 &     $-0.38$       &     $-0.33$                  \\
 Ti(0001) &  $-196$~~~~~~~~  &    $21.82$                 &     $-0.87$       &     $-0.70$                  \\
  Cu(111) &  $-277$~$(-190)$ &    $~9.14$                 &     $-0.96$       &     $-0.83$                  \\
  Pt(111) &  $-367$~$(-320)$ &    $49.09$                 &     $-1.36$       &     $-0.98$                  \\ 
  Pd(111) &  $-453$~$(-360)$ &    $51.22$                 &     $-1.44$       &     $-1.28$                  \\
  \hline\hline
          &\multicolumn{2}{c|}{$d_{\rm Xe-metal}$~(\AA)}  &\multicolumn{2}{c}{substrate rumpling~(\AA)     } \\  
          &       on-top     &      fcc                   &\multicolumn{2}{c}{       on-top                } \\  \hline\hline
 Mg(0001) & $3.80$~~~~~~~~   &    $3.84$                  &\multicolumn{2}{c}{$~0.00$~~~~~~~~~~~~~~~~~~~~~ } \\
  Al(111) & $3.69$~~~~~~~~   &    $3.72$                  &\multicolumn{2}{c}{$-0.01$~~~~~~~~~~~~~~~~~~~~~ } \\
 Ti(0001) & $3.56$~~~~~~~~   &    $3.66$                  &\multicolumn{2}{c}{$+0.06$~~~~~~~~~~~~~~~~~~~~~ } \\
  Cu(111) & $3.25$~$(3.60)$  &    $3.31$                  &\multicolumn{2}{c}{$+0.03$~$(-0.01 \pm 0.02)$   } \\
  Pt(111) & $3.07$~$(3.40)$  &    $3.19$                  &\multicolumn{2}{c}{$+0.04$~$(-0.01 \pm 0.02)$   } \\  
  Pd(111) & $2.85$~$(3.07)$  &    $2.86$                  &\multicolumn{2}{c}{$+0.05$~$(-0.02 \pm 0.03)$   } \\
\end{tabular}
\end{table}

Our calculations show: $(i)$ The on-top site is energetically favored for
{\em all} studied systems in the $\sqrt{3}$ structure. $(ii)$ Calculations
for low Xe coverage, $\Theta_{\rm Xe}$ $=$ $1/4$ and $1/9$, show that also
in these cases Xe atoms prefer the on-top site (e.g., for
Xe/Pd(111) and Xe/Pt(111) the on-top site is favored over the fcc site by
$20.59$~meV and $57.87$~meV, respectively, at $\Theta_{\rm Xe}$ $=$ $1/9$
using the LDA). Thus, lateral Xe--Xe interactions are {\em not} responsible
for the on-top Xe site preference. $(iii)$ Spin-orbit coupling corrections
were included for the valence-states using the second variational
method~\cite{Blaha-1990};
this yields a change 
in the energy difference between on-top
and fcc sites of $\leq$~2~meV, and the
geometrical parameters are unchanged. $(iv)$ The energy difference between
the on-top and fcc sites is smaller in the GGA than in the LDA (e.g., for
Xe/Pd(111) the on-top site is favorable by $13.48$~meV (GGA) and $51.22$~meV
(LDA), while for Xe/Pt(111) the on-top site is favorable by $4.80$~meV (GGA)
and $49.09$~meV (LDA)). Furthermore, the GGA predicts a weaker adsorption energy
than LDA (e.g., $E_{\rm ad}^{\rm Xe/Pd(111)}$ $=$ $-76$~meV (GGA)
and $-453$~meV (LDA), $E_{\rm ad}^{\rm Xe/Pt(111)}$ $=$ $-82$~meV (GGA) and
$-367$~meV (LDA)). However, both functionals gives the {\em same} 
qualitative picture for the
interaction between Xe atoms and metal surfaces. In view of these four points
we limit our discussion in the following to the LDA results, neglecting
spin-orbit coupling corrections for the valence states, and discussing only
the $\sqrt{3}$ structure. All our results,
including convergence tests,
will be presented in detail in a
forthcoming publication~\cite{Dasilva}. 

\begin{figure}[t]
 \pspicture(5,5.5)
 \rput[c](4.3,3.0){
 \psfig{figure=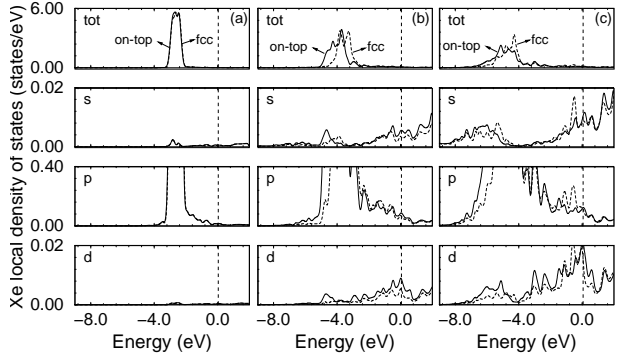,height=5.0cm}}
 \endpspicture
\caption{Local density of states for Xe on Pt(111) 
for different vertical positions of Xe, $Z$: 
(a) $Z$ $=$ $4.53$~{\AA}; (b) $Z$ $=$
$3.07$~{\AA} (equilibrium Xe position); (c) $Z$ $=$ $2.70$~{\AA}.
}
\end{figure}

Figures 1a and 1b show respectively the relative binding energy (with respect 
to the on-top site geometry) and the adsorption induced work-function change 
for Xe on Pt(111). It can be seen that the 
on-top site is  energetically favorable and Xe adsorption 
induces a decrease in the work-function (which is {\em greater} for the 
on-top site), in agreement with experiment. The adsorption energy,
the relative energy difference 
for Xe in the on-top and fcc sites, as well as the work-function 
change for all the systems are summarized in Tab.~I, where
it can be seen that this same 
behavior is exhibited.
For the cases where there are experimental values of the adsorption
energy with which to compare (Xe/Cu, Xe/Pd, and Xe/Pt), 
the agreement between theory and experiment is quite good. 

In Tab.~I we also list the important structural data. The DFT-LDA bond
lengths are about 10\% shorter than those obtained by
LEED~\cite{Seyller-1998}, but in view of the fact that the adsorption energy
is weak in physisorption systems and our theory does not include vibrational
contributions, the sign and size of the deviation is as expected. The {\em
difference} in bondlengths, however, for the Xe/Cu, Xe/Pd, and Xe/Pt systems,
for which comparison is possible, are in excellent agreement between theory
and LEED analyses. There is one deviation between the calculations and
LEED analyses and that is in the direction of displacement of the atom
below the Xe atom.
The experimental values lie however within the given error margins. 
Particularly interesting is that the vertical distance between the Xe adatom 
and the topmost metal layer is {\em smaller} for the on-top site as compared 
to the fcc site for {\it all} the studied cases (for Xe/Pd it is almost 
equal). 
This is in {\em contrast} to the calculations of 
Ref.~\onlinecite{Muller-1990}, and in contrast to what might be intuitively 
expected. We will come back to this point below.

Figures 2a, 2b, and 2c show the local density of states (LDOS) for Xe on Pt(111)
for three distances of Xe above the surface. It can be 
seen that 
there is a broadening of the Xe states as the Xe adatom approaches the 
surface,
which leads to a partial occupation of the 
previously unoccupied Xe $6s$- and $5d$-states (mainly $5d_{\rm z^2}$-state), 
and a partial depopulation of the Xe $5p$-states (mainly the Xe 
$5p_{\rm z}$-state). 
The same behavior in the LDOS 
is identified for Xe on {\it all} 
the studied metal surfaces. 
We note that in Ref.~\onlinecite{eigler}, occupation of the Xe-$6s$
resonance was attributed to being the origin of the Fermi level
LDOS which renders Xe visible on Ni(110) in scanning tunneling
microscopy experiments. This is consisent with the present results.
\begin{figure}[t]
 \pspicture(5,5.5)
 \rput[c](4.3,3.0){
 \psfig{figure=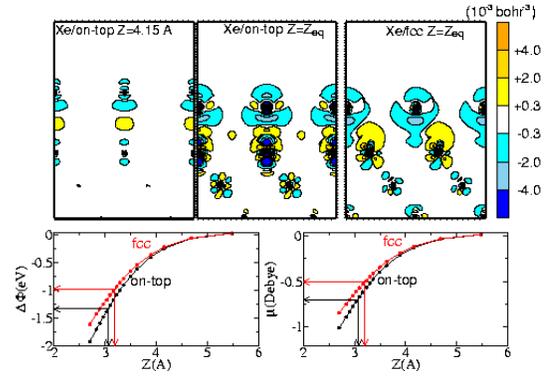,height=5.0cm}}
 \endpspicture
\caption{Upper panels:
Difference electron density, $n^{\Delta}({\bf r}) = n^{\rm Xe/Pt}({\bf r})$ 
$-$ $n^{\rm Pt}({\bf r})$ $-$ $n^{\rm Xe}({\bf r})$, along the 
$(11\bar{2})$ plane for Xe on Pt(111)
in the on-top site at 4.15~\AA\, above the equilibrium
position (left), in the on-top equilibrium site (Z=Z$_{\rm eq}$) (middle),
and in the fcc-hollow equilibrium site (right).
Lower panels: Work-function change, $\Delta \Phi$, (left) and dipole
moment, $\mu$, (right) versus distance of Xe from the surface. The points
are the DFT-LDA values.   
}
\label{diff-rho}
\end{figure}        

We now turn to consider the change in electron density due to adsorption, 
i.e., the difference between the density of the interacting system, 
$n^{\rm Xe/Pt}$, and the sum of the densities of the clean surface, 
$n^{\rm Pt}$, and a free Xe layer, $n^{\rm Xe}$. 
We find that a polarization of the Xe atom sets in when it is already far
from the surface, where the negative end of the dipole points towards
the surface, i.e., the electron density is enhanced closer to the 
metal surface
(see Fig.~3, upper left).  This is due to the general property that
the potential in front of a metal is attractive for electrons;
specifically, the exchange-correlation potential is more attractive
on the metal side of the Xe atom
than on the vacuum side. At this large distance where the Xe
dipole moment starts to build up, the electrostatic and
covalent contributions are small, if present at all.
This behavior is consistent with the pioneering works of
Lang~\cite{lang} and Antoniewicz~\cite{anton}.
A corresponding polarization occurs in the surface,
which can be described as the image of the Xe dipole.  The build-up
of these dipoles happens simultaneously (and continuously
with decreasing distance from the surface) and are related 
to the partial population and depopulation of the Xe 
states (cf. Fig.~2). In Fig.~3, 
the work-function change and dipole moment is shown
as a function of distance, Z. For all Z, the values for the on-top
site are greater and the variation with distance is smooth and continuous.
From the electron density difference distributions at the equilibrium
positions (Z=Z$_{\rm eq}$)
depletions of charge
are seen at the Xe atom (mainly $5p_{\rm z}$) and
at the metal atom closest to Xe, i.e., mainly in
the Pt $d_{\rm z^2}$-states, while an 
increase occurs at the surface, between these regions of depletion.  
Similar results are found for all the other 
studied transition metal surfaces. 

The differences in the LDOS and $\Delta n({\bf r})$ distributions 
between on-top and hollow sites described above, show     
that the Pauli repulsion 
and the attractive interaction is {\em not} the same at each site.
In order to investigate this aspect in more detail, we 
calculated the binding energy of Xe on the metal surfaces as a function of 
vertical distance. The results for Xe/Pt are 
shown in Fig.~4 (left).  
At all distances above the equilibrium position, the on-top 
site has lower energy. 
From these results,
we obtain Xe vibrational frequencies of $E_{\rm vib}$ $=$ $3.79$~meV 
(on-top) and $4.17$~meV (fcc).
Reported experimental values are 3.70~meV and 3.80~meV,~\cite{Zeppenfeld-LB}
showing excellent agreement with the calculated (on-top) value.

In Fig. 4 (right), we decomposed the binding energy curves into a repulsive 
and attractive potential using a standard empirical formula: 
$E(Z)=A\, exp(-BZ) - C(Z-Z_{0})^{-3}$~\cite{Bruch-1997}, where $Z_{0}$ is
the position of the image plane. 
We note that the two right-most DFT-LDA points in Fig.~4 (left) were
not included in the fit since DFT results are most reliable 
near the minimum in the potential well.
It can be seen 
quite clearly that the Pauli repulsion is different 
for the two sites and is {\em weakest} for the on-top site. We also used 
other expressions for the repulsive and attractive terms, and the conclusion 
was always the same.
We furthermore decomposed the DFT-total energy into kinetic, Coulomb, and
exchange-correlation energies. The results suggest that the repulsion
due to the kinetic energy is smaller for Xe in the on-top site.
We find the same behavior as exhibited in Fig.~4 
for {\em all} the rare-gas atom--metal systems investigated.

\begin{figure}[t]
 \pspicture(5,5.5)
 \rput[c](4.3,2.6){
 \psfig{figure=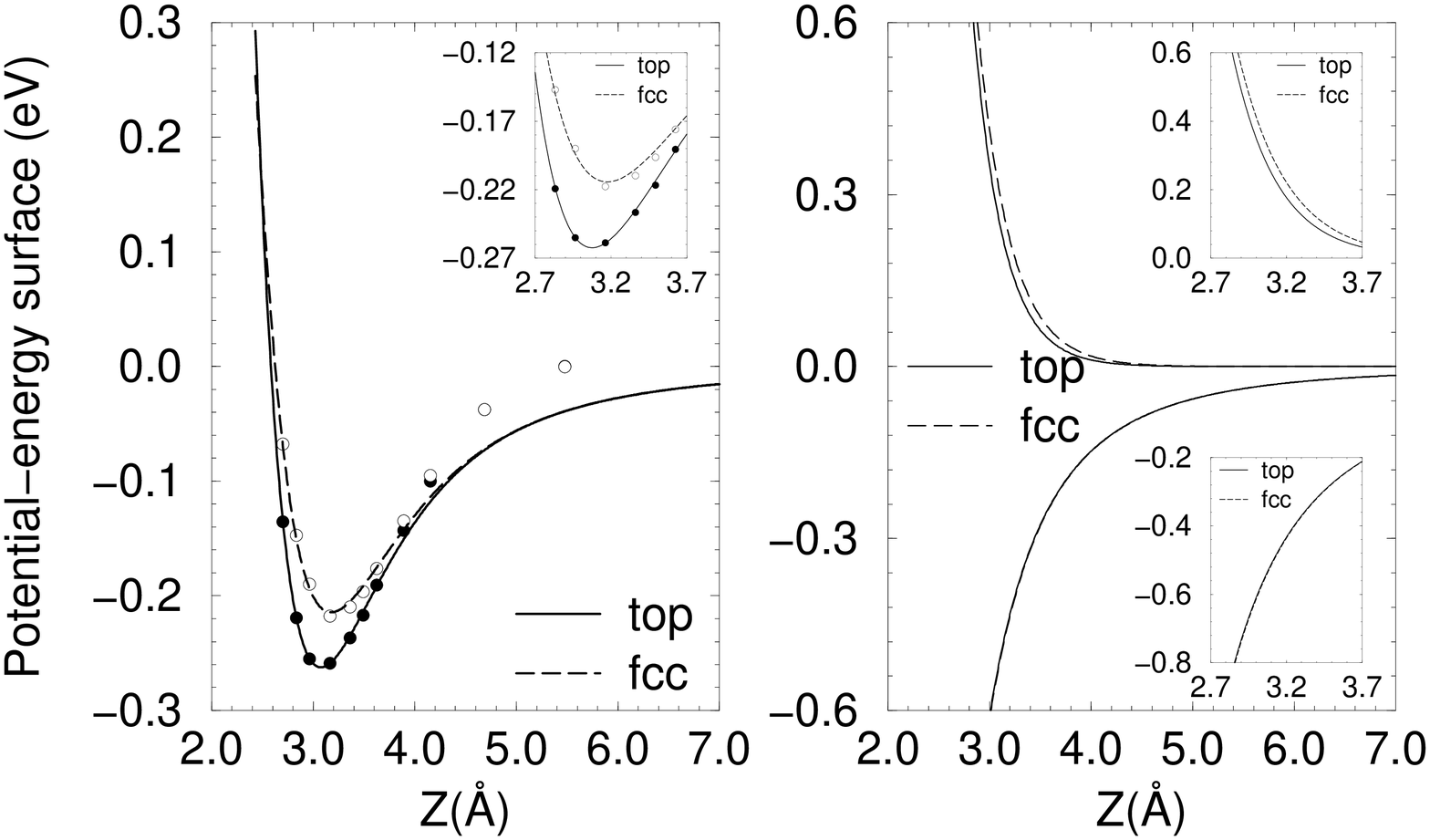,height=5.2cm}} 
 \endpspicture
\caption{Left panel: Calculated binding energies of Xe on Pt(111) 
for the on-top and fcc 
sites as a function of distance from the surface. 
The filled and open circles indicate DFT-LDA data. 
Right panel: decomposition into repulsive and attractive contributions 
(see text). $Z=0$ is the center of the top layer.}
\label{pauli}
\end{figure}        
The reason why, that for Xe at the on-top site, the Pauli repulsion is weaker,
the equilibrium distance closer to the surface, 
and the dipole moment greater, is
explained as follows:
On analyzing the reactivity index of the clean Pt surface~\cite{wilke}, we find
that the electronic character at the on-top region is donor-like
and at the hollow region it is acceptor like. Thus, for a
negative charge, at not too close distance (or for a dipole
with the negative end pointing toward the surface), the
on-top region can easily screen this perturbation
by transferring electron density from the on-top region
(i.e., from $d_{z^2}$ and $p_z$ orbitals) to the interstitial
region. Therefore, at intermediate distances the Xe atom
becomes polarized more strongly at the on-top geometry,
because the screening charge at the metal can build up
more efficiently. This is why the dipole is greater.
Similarily, at close distance, when
wavefunctions start to overlap and Pauli repulsion sets in,
the substrate can reduce this repulsion more efficiently
at the on-top site by transferring $d_{z^2}$ and $p_z$
electrons to $s$-states. Hence Xe can get closer to the surface.

In summary, from DFT calculations
we determined that Xe adatoms on the
free-electron-like, early and late transition metals, as well as noble
metals, Mg(0001), Al(111), Ti(0001), 
Cu(111), Pd(111), and Pt(111), prefer low coordination on-top sites. The 
qualitative results are the same for LDA and GGA functionals. 
We find that the dominant mechanisms 
at play in these systems are polarization and a site-dependent 
Pauli repulsion which is weaker for the on-top site, thus stabilizing
on-top adsorption. 
Our 
recent investigations for the smaller rare-gas atoms, e.g., Ar and Kr on 
Pd(111) exhibit the same behavior and basic mechanisms,
and we anticipate that the present findings may also 
be generally valid for other rare-gas/metal systems,
thus presenting a new picture of rare-gas adsorption, where
the on-top site is the rule rather than the exception.


\vspace{-0.5cm}
\begin{references}
\vspace{-1.5cm}  

\bibitem{Bruch-1997} L. W. Bruch, M. W. Cole, and E. Zaremba, in {\em 
Physical Adsorption: Forces and Phenomena}, Oxford Science Publications (1997).

\bibitem{Zeppenfeld-LB} P. Zeppenfeld, in {\em Physics of covered solid 
surfaces I. Adsorbed layers on surfaces}, Landolt-B\"{o}rnstein (to be 
published), see reference therein. 

\bibitem{Petersen-1996} M. Petersen et al., Phys. Rev. Lett. {\bf 76}, 995 
(1996).

\bibitem{Gottlieb-1990} J. M. Gottlieb, Phys. Rev. B {\bf 42}, 5377 (1990).

\bibitem{Kern-1987} K. Kern et al., Solid State Commun. {\bf 62}, 391 (1987).

\bibitem{Zeppenfeld-1988} P. Zeppenfeld et al., Phys. Rev. B {\bf 38}, 3918 
(1988).

\bibitem{Potthoff-1995} M. Potthoff et al., Surf. Sci {\bf 322}, 193 (1995).

\bibitem{Hilgers-1995} G. Hilgers, et al., Surf. Sci. {\bf 322}, 207 (1995). 


\bibitem{Seyller-1998} Th. Seyller et al., Chem. Phys. Lett. {\bf 291}, 567 
(1998); Phys. Rev. B {\bf 60}, 11084 (1999).


\bibitem{Muller-1990} J. E. M\"{u}ller, Phys. Rev. Lett. {\bf 65}, 3021 (1990).

\bibitem{Betancourt-2000} A. E. Betancourt and D. M. Bird, J. Phys. Condens. 
Mat. {\bf 12}, 7077 (2000). 

\bibitem{Clarke-2001} S. Clarke, G. Bihlmayer, and S. Bl\"{u}gel, Phys. Rev. 
B {\bf 63}, 85416, (2001). 

\bibitem{bagus} P. S. Bagus et al. Phys. Rev. Lett. {\bf 89}, 096104
(2002).


\bibitem{Blaha-1990} P. Blaha, K. Schwarz, and J. Luitz, WIEN97, A 
Full-Potential Linearized Augmented Plane Wave Package for Calculating 
Crystal Properties, (Karlheinz Schwarz, Techn. Univ. Wien, Vienna 1999). 
ISBN 3-9501031-0-4. 

\bibitem{Petersen-2000} M. Petersen et al., Comput. Phys. Commun. {\bf 126}, 
294 (2000); R. Dohmen et al., Comput. Sci. \& Engin. {\bf 3}, 18 (2001). 

\bibitem{Perdew-1992} J. P. Perdew and Y. Wang, Phys. Rev. B {\bf 45}, 13244 
(1992).

\bibitem{Perdew-1996} J. P. Perdew, S. Burke, and M. Ernzerhof, Phys. Rev. 
Lett. {\bf 77}, 3865 (1996).

\bibitem{Dasilva} J. L. F. Da Silva, C. Stampfl, and M. Scheffler, in preparation.

\bibitem{eigler} D. M. Eigler et al. Phys. Rev. Lett. {\bf 66}, 1189
(1991).  
\bibitem{lang} N. D. Lang, Phys. Rev. Lett. {\bf 46}, 842 (1981).
\bibitem{anton} P. R. Antoniewicz, Phys. Rev. Lett. {\bf 32}, 1424 (1974).
\bibitem{wilke} S. Wilke, M. H. Cohen, and M. Scheffler, 
Phys. Rev. Lett. {\bf 77}, 1560 (1996).

\end{references}
\end{document}